\documentclass[aps,prd,twocolumn,superscriptaddress,showpacs,floatfix,handout,10pt]{revtex4-1}

\usepackage{amsmath,amssymb,mathtools}
\usepackage{color}
\usepackage{graphicx}
\usepackage{subfigure}
\usepackage[colorlinks=true,linkcolor=red,citecolor=blue,urlcolor=blue,bookmarks]{hyperref}
\usepackage{multirow,makecell}
\usepackage{textcomp}
\usepackage{wasysym}
\usepackage{ulem}

\usepackage{verbatim}

\usepackage[utf8]{inputenc}
\usepackage[T1]{fontenc}

\usepackage[text={17.1cm,24.6cm},centering]{geometry} 

\def \be {\begin{equation}}
\def \ee {\end{equation}}
\def \ba {\begin{array}}
\def \ea {\end{array}}
\def \bea {\begin{eqnarray}}
\def \eea {\end{eqnarray}}

\def \d {\delta}
\def \D {\Delta}

\def \ii {\mathrm{i}}

\def \and {{~\textrm{and}~}}

\begin{document}

\title{
\textbf{Single-eigenstate test of eigenstate thermalization hypothesis via\\perturbed eigenstate quench}
}


\author{Zhouhao Guo}
\affiliation{Center for Joint Quantum Studies and Department of Physics, School of Science,
Tianjin University, 135 Yaguan Road, Tianjin 300350, China}

\author{Jiaju Zhang}
\email{jiajuzhang@tju.eud.cn}
\affiliation{Center for Joint Quantum Studies and Department of Physics, School of Science,
Tianjin University, 135 Yaguan Road, Tianjin 300350, China}


\begin{abstract}

  We propose and numerically validate an efficient single-eigenstate diagnostic for the eigenstate thermalization hypothesis (ETH) based on a perturbed eigenstate quench protocol. By introducing a weak random perturbation to an energy eigenstate to break its stationarity, we characterize the time-averaged subsystem evolution speed as a function of the subsystem-to-total system size ratio. The diagnostic relies on a robust qualitative distinction: eigenstates satisfying ETH exhibit an S-shaped curve with a clear inflection point near half the system size, while ETH-violating eigenstates display a convex J-shaped profile. We benchmark the criterion across paradigmatic one-dimensional spin chains covering chaotic, integrable, many-body localized, and quantum many-body scar regimes, obtaining full agreement with established thermalization phenomenology. Our method circumvents the need for explicit thermal ensemble construction, providing a robust, experimentally feasible probe of eigenstate thermalization at the single-eigenstate level.

\end{abstract}

\maketitle



\tableofcontents

\section{Introduction}\label{sectionIntroduction}

Understanding the microscopic origin of thermalization in isolated quantum many-body systems has long been a fundamental and intriguing problem at the intersection of quantum statistical mechanics and condensed matter physics. A cornerstone theoretical framework for addressing this issue is the eigenstate thermalization hypothesis (ETH) \cite{Deutsch:1991msp,Srednicki:1994mfb,Rigol:2007mja}, which provides a mechanism for how closed quantum systems approach thermal equilibrium. The core statement of ETH is that, for generic chaotic many-body Hamiltonians, the expectation value of any local observable computed in a single energy eigenstate matches the prediction from the corresponding thermal ensemble at the same energy density. In this picture, thermal properties are already encoded within individual eigenstates, such that unitary dynamics naturally drives generic initial states toward thermal equilibrium.

Not all quantum many-body systems obey ETH and exhibit conventional thermalization. Integrable systems, for instance, possess an extensive set of local conserved quantities that constrain their dynamics; as a consequence, they generically relax to generalized Gibbs ensembles (GGEs) rather than standard thermal states \cite{Rigol:2006jrd}. For clarity, we note that the term ``thermal state'' as used throughout this paper encompasses both conventional thermal ensembles and GGEs. A class of non-thermalizing systems emerges in the presence of strong disorder, which can induce a many-body localized (MBL) phase. In the MBL phase, the emergence of a complete set of quasi-local conserved quantities completely suppresses ergodic dynamics and prevents thermalization even at infinite time \cite{Gornyi:2005mbj,Basko:2006fmp,Nandkishore:2014kca,Alet:2017dwx,Abanin:2018yrt,Sierant:2024khi}. Beyond the extreme cases of full ergodicity and full localization, an intermediate regime of weak ergodicity breaking has been discovered, in which the majority of eigenstates satisfy ETH while a small, sparse set of eigenstates violates it. These exceptional non-thermal eigenstates are known as quantum many-body scar states, and they represent a novel paradigm for ergodicity breaking \cite{Bernien:2017ubn,Turner:2017fxc,Serbyn:2020wys,Moudgalya:2021xlu,Chandran:2022jtd}.

Given the central status of ETH in quantum thermalization, developing efficient and reliable diagnostics to verify ETH at the level of individual eigenstates is of significant importance. Several testing methods have been proposed in the literature, each with distinct advantages and limitations. The most direct approach, rooted in the original ETH formulation, is to examine the matrix elements of local operators in the energy eigenbasis. However, this method is computationally cumbersome in practice, as it requires sampling and analyzing numerous local operators to draw statistically robust conclusions. An alternative single-eigenstate diagnostic is to compute the trace distance between the reduced density matrix (RDM) of an energy eigenstate and that of the corresponding thermal state, and study its scaling with subsystem size \cite{Garrison:2015lva,Dymarsky:2016ntg}. While this approach targets individual eigenstates, the explicit construction of the matching thermal ensemble can be difficult. This drawback has motivated the search for alternative ETH probes that avoid the explicit construction of thermal states.

Several recent proposals aim to bypass this computational cost, but they come with other restrictions. For example, the average subsystem trace distance between adjacent eigenstates in the energy spectrum has been employed as an indicator of ETH \cite{Khasseh:2023kxw}, but this quantity characterizes the global properties of the spectrum and cannot be applied to diagnose a single eigenstate. Another work uses the subsystem evolution speed as a probe of stationary states after a quantum quench \cite{Zhang:2024qya}; however, since energy eigenstates are stationary by definition, neither the full eigenstate nor its RDM undergoes any evolution, so this method cannot be directly used to test ETH for individual eigenstates.

To overcome the limitations outlined above, in this paper we propose a novel single-eigenstate diagnostic for ETH that does not require explicit construction of thermal ensembles. Instead of directly comparing an eigenstate to a thermal state via trace distance, we study the subsystem evolution speed of an energy eigenstate subjected to a small random perturbation. Our central idea builds on the ETH prediction that in a thermalizing eigenstate, a subsystem smaller than half the total system size exhibits thermal behavior, while a subsystem larger than half does not, due to the distinct entanglement structures of eigenstates and thermal states \cite{Garrison:2015lva}. Under a random perturbation, small thermalizing subsystems and large non-thermalizing subsystems respond differently, leading to a characteristic transition in the subsystem evolution speed near half the system size. For eigenstates that violate ETH, by contrast, this thermal-like entanglement structure is absent, and thus no such transition occurs.

Specifically, we analyze the curve of subsystem evolution speed as a function of $x$, the ratio of subsystem size to total system size. The evolution speed is always monotonically increasing with $x$. For eigenstates satisfying ETH, the curve adopts an S-shaped profile: it is convex for $x<1/2$ and concave for $x>1/2$, with a transition point near $x=1/2$. For eigenstates breaking ETH, however, the curve follows a J-shaped profile, remaining convex throughout the range $0<x<1$ with no inflection point at half system size. This qualitative difference in curve shape provides a clear and efficient signature for distinguishing ETH-satisfying from ETH-violating eigenstates.

The remainder of this paper is structured as follows. We first introduce the perturbed eigenstate quench protocol and give the precise definition of the subsystem evolution speed. We then apply our diagnostic to several representative spin chain models spanning distinct thermalization regimes: chaotic and integrable transverse-field Ising chains, disordered XXZ chains in both the ergodic and MBL regimes, and a spin-1 XX chain that hosts quantum many-body scar states. We demonstrate that our method yields distinct, characteristic behaviors for ETH-satisfying and ETH-breaking eigenstates across all these model systems, confirming its utility as a robust ETH diagnostic at the single-eigenstate level.

\section{Perturbed eigenstate quench}

We now detail the numerical implementation of our proposed single-eigenstate ETH diagnostic. For a given many-body Hamiltonian $H$, we first obtain its complete set of energy eigenstates $\{|\psi_i\rangle\}$ via exact diagonalization, where the index $i$ labels eigenstates ordered by their corresponding energy eigenvalues $E_i$. To probe the dynamical signatures encoded in an individual eigenstate $|\psi_i\rangle$, we introduce a weak random perturbation to construct a non-stationary initial state that remains close to $|\psi_i\rangle$.
Specifically, the perturbed initial state takes the form
\begin{equation}
|\psi(0)\rangle = \frac{1}{\mathcal{N}} ( |\psi_i\rangle + \epsilon |\psi_{\rm random}\rangle ),
\end{equation}
where $\mathcal{N}$ is the normalization factor, $\epsilon$ is a small parameter controlling the perturbation strength, and $|\psi_{\rm random}\rangle$ is a pure state sampled uniformly from the full Hilbert space according to the Haar measure. The small perturbation strength ensures that the initial state inherits the core structural properties of the target eigenstate $|\psi_i\rangle$, while the admixture of a random component breaks the stationary condition and induces nontrivial time evolution. In all numerical calculations presented in this work, we fix the perturbation strength at $\epsilon = 10^{-3}$.
Under unitary evolution governed by the original Hamiltonian $H$, the time-dependent state at time $t$ is given by
\begin{equation}
|\psi(t)\rangle = e^{-\mathrm{i} H t} |\psi(0)\rangle.
\end{equation}

To characterize the local dynamical behavior, we focus on the RDM of a subsystem $A$, obtained by tracing out the degrees of freedom of its complement subsystem $B$
\begin{equation}
\rho_A(t) = \mathrm{tr}_B ( |\psi(t)\rangle \langle \psi(t)| ).
\end{equation}
The time evolution of $\rho_A(t)$ captures all dynamical information of local observables supported on subsystem $A$, making it a natural quantity for investigating thermalization properties at the subsystem level.
We quantify the rate of change of the subsystem state via the subsystem evolution speed, defined as the trace distance between RDMs at infinitesimally separated times \cite{Zhang:2024qya}
\begin{equation}
v_A(t) = \lim_{\d t \to 0} \frac{D\left( \rho_A(t+\delta t), \rho_A(t) \right)}{\delta t}.
\end{equation}
Here, $D(\rho, \sigma) = \frac{1}{2} \| \rho - \sigma \|_1$ denotes the trace distance between two density matrices $\rho$ and $\sigma$, with $\| \cdot \|_1$ denoting the Schatten 1-norm (trace norm). As a standard measure of quantum state distinguishability, the trace distance provides a well-motivated metric for quantifying state evolution. In the limit $\delta t \to 0$, $v_A(t)$ converges to the instantaneous time derivative of the trace distance. For our numerical simulations, we choose a sufficiently small time step $\delta t = 10^{-6}$ to well approximate the continuous-time limit.

To obtain statistically stable results and suppress transient oscillations and numerical noise, we uniformly sample $N$ time points ${t_n}$ within a properly chosen time interval and compute the time-averaged subsystem evolution speed
\begin{equation}
\langle v_A(t) \rangle = \frac{1}{N} \sum_{n=1}^N v_A(t_n).
\end{equation}
This time-averaged quantity characterizes the typical rate at which the reduced density matrix of subsystem $A$ evolves, and serves as the core observable in our diagnostic.

By systematically varying the size $\ell$ of subsystem $A$ and computing $\langle v_A \rangle$ for each subsystem size, we obtain the dependence of the averaged evolution speed on the subsystem-to-total-system size ratio $x = \ell / L$, where $L$ is the total system size. Since the trace distance is monotonically increasing with the dimension of the subsystem \cite{Nielsen:2010oan}, the subsystem evolution speed is also a monotonically increasing function of $x$ in general. Crucially, the qualitative shape of the $\langle v_A(t) \rangle$ versus $x$ curve exhibits distinct features for ETH-satisfying and ETH-violating eigenstates, providing a clear signature to identify ergodicity breaking at the single-eigenstate level. Since our analysis focuses on the qualitative curve profiles, we rescale the time-averaged subsystem evolution speed as $\langle v_A(t) \rangle/v_{\rm whole}$, where $v_{\rm whole}$ denotes the evolution speed of the full system.

\section{Chaotic and integrable Ising chains}

We begin our numerical investigation with the one-dimensional spin-1/2 Ising chain with both transverse and longitudinal fields. The Hamiltonian, which includes nearest-neighbor spin-exchange interactions and on-site magnetic fields, takes the form
\begin{equation}
H = -\frac{1}{2} \sum_{j=1}^{L} (
\sigma_j^x \sigma_{j+1}^x + h_x \sigma_j^x + h_z \sigma_j^z
),
\end{equation}
where $\sigma_j^\mu$ ($\mu=x,y,z$) are the Pauli matrices acting on lattice site $j$. We impose periodic boundary conditions such that $\sigma_{L+1}^x = \sigma_1^x$.

This model exhibits markedly different integrability structures as its field parameters are tuned. When the longitudinal field vanishes ($h_x = 0$), the model can be exactly mapped to a system of noninteracting fermions via successive Jordan-Wigner, Fourier, and Bogoliubov transformations and is therefore fully integrable \cite{Lieb:1961fr,Katsura:1962hqz,Pfeuty:1970ayt}. In the integrable regime, relaxation dynamics are constrained by an extensive set of conserved quantities, and post-quench states thermalize to the GGE \cite{Calabrese:2011vdk,Cassidy:2011smq}. By contrast, when both $h_x$ and $h_z$ are nonzero, the additional longitudinal field induces effective nonlocal interactions in the fermionic picture that break integrability, drive the system into a quantum chaotic regime, and cause post-quench states to thermalize to the conventional canonical thermal state \cite{Banuls:2010fkc,Kim:2014jfl}.

\begin{figure}[tp]
  \centering
  \includegraphics[width=0.23\textwidth]{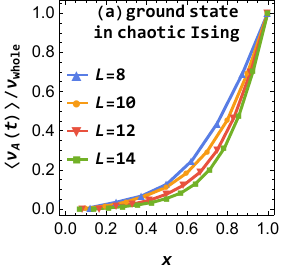}
  \includegraphics[width=0.23\textwidth]{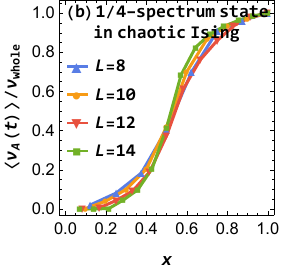}\\
  \includegraphics[width=0.23\textwidth]{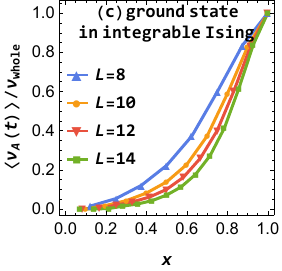}
  \includegraphics[width=0.23\textwidth]{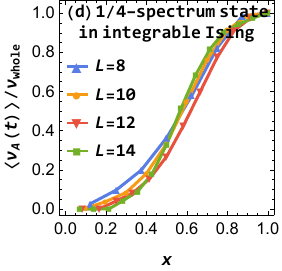}
  \caption{Rescaled time-averaged subsystem evolution speed after a perturbed eigenstate quench in chaotic and integrable Ising chains. \textit{Top panels}: Chaotic Ising chain with parameters $h_x = \frac{\sqrt{3}}{2}$ and $h_z = \sqrt{2}$. \textit{Bottom panels}: Integrable Ising chain with parameters $h_x = 0$ and $h_z = \sqrt{2}$. \textit{Left panels}: Results for the ground state, which violates ETH. \textit{Right panels}: Results for the eigenstate at the 1/4 position of the energy spectrum, which satisfies ETH. The time average is taken over the interval $t\in[2L,4L]$.}
  \label{FigureIsing}
\end{figure}

The integrable Ising chain is solved exactly following Refs. \cite{Lieb:1961fr,Katsura:1962hqz,Pfeuty:1970ayt}, while the chaotic Ising chain is solved numerically via block diagonalization as described in Ref. \cite{Sandvik:2010lkj}. We compute the time-averaged subsystem evolution speed following a perturbed eigenstate quench for both the chaotic and integrable Ising chains, with results presented in Fig.~\ref{FigureIsing}. We examine two representative eigenstates: the ground state at the bottom of the spectrum, and a bulk eigenstate located at the 1/4 position of the full energy spectrum.

The ground state violates ETH in both the chaotic and integrable chains. States near the lower and upper edges of the spectrum similarly break ETH, as the low density of states at spectral edges prevents the emergence of typical thermal behavior. For the bulk eigenstate at the 1/4 spectral position, by contrast, ETH is satisfied in both regimes, though with distinct physical meanings: in the chaotic chain it corresponds to conventional thermalization to a canonical ensemble, while in the integrable chain it corresponds to generalized thermalization to a GGE.

Our numerical results establish a clear qualitative criterion to distinguish eigenstates that obey the ETH from those that violate it. For ETH-satisfying eigenstates, the time-averaged subsystem evolution speed as a function of the subsystem-to-total size ratio $x$ adopts a characteristic S-shaped profile: the curve is convex for $x<1/2$, becomes concave for $x>1/2$, and features a well-defined inflection point near $x=1/2$. For $x<1/2$, the subsystem is smaller than half the total system size and exhibits thermal behavior, making it more robust against weak perturbations and yielding a comparatively low evolution speed. Conversely, for $x>1/2$, the subsystem exceeds half the system size and deviates from thermal character, rendering it more susceptible to perturbations and resulting in a higher evolution speed. By contrast, ETH-violating eigenstates display a J-shaped profile that remains convex across the full range $0<x<1$, with no inflection point at the half-system-size scale. This robust structural distinction provides a direct, visually unambiguous signature for identifying ergodicity breaking at the single-eigenstate level.

\section{Disordered XXZ chain in chaotic and MBL regimes}

We then consider a disordered spin-$1/2$ XXZ chain of length $L$, whose Hamiltonian takes the form
\begin{equation}
H = \sum_{j=1}^{L} \Big[
\frac{1}{4} ( \sigma^x_j \sigma^x_{j+1} + \sigma^y_j \sigma^y_{j+1} + \Delta \sigma^z_j \sigma^z_{j+1} )
+ \frac{1}{2} h_j^z \sigma^z_j
\Big].
\end{equation}
Periodic boundary conditions are imposed such that $\sigma^\alpha_{L+1} = \sigma^\alpha_1$ for all spin components $\alpha=x,y,z$. The on-site longitudinal fields $\{h_j^z\}$ are independent random variables uniformly distributed over the interval $[-W, W]$, where the parameter $W$ quantifies the overall strength of disorder in the system.

The introduction of quenched random fields explicitly breaks the translational symmetry of the clean integrable XXZ chain, meaning that momentum is no longer a conserved quantity. Concurrently, the site-dependent disorder potentials destroy the integrable structure of the homogeneous model, giving rise to a generic nonintegrable interacting many-body system. A paradigmatic feature of this disordered spin chain is the emergence of MBL when disorder is sufficiently strong \cite{Pal:2010ezp,Luitz:2015cep}. As the disorder strength $W$ is tuned, the system undergoes a phase transition between two qualitatively distinct dynamical regimes. In the weak-disorder (thermal/chaotic) regime, the system conforms to the ETH: typical eigenstates exhibit volume-law entanglement, and the long-time expectation values of local observables converge to their thermal equilibrium values. With increasing disorder strength, the system gradually deviates from thermalizing behavior, and eventually transitions into the MBL phase, where ergodicity is broken and the system fails to thermalize even at long times.

\begin{figure}[tp]
  \centering
  \includegraphics[width=0.23\textwidth]{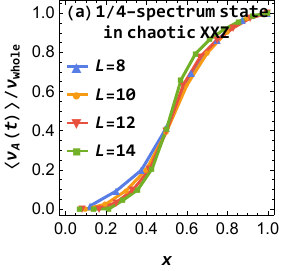}
  \includegraphics[width=0.23\textwidth]{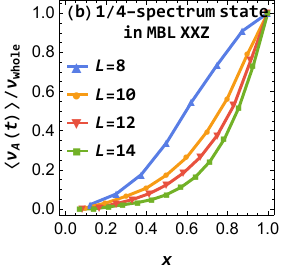}
  \caption{Rescaled time-averaged subsystem evolution speed after a perturbed eigenstates quench in a disordered XXZ chain, shown for both the chaotic and MBL regimes. \textit{Left}: Chaotic regime with disorder strength $W = \sqrt{2}$. \textit{Right}: MBL regime with disorder strength $W = \sqrt{43}$. We have taken $\D=1$ for both panels. The time average is computed over the time window $t\in[2L,4L]$.}
  \label{FigureXXZ}
\end{figure}

Figure~\ref{FigureXXZ} presents the rescaled time-averaged subsystem evolution speeds following a quench from perturbed eigenstates, for the disordered XXZ chain in both the chaotic and MBL regimes. From the profile of the curves, we conclude that the 1/4-spectrum eigenstate satisfies ETH in the chaotic regime, but violates ETH in the MBL regime, consistent with the established phenomenology of the two phases.

\section{Spin-$1$ XX chain}

Finally, we study a one-dimensional spin-$1$ XX chain of length $L$, governed by the Hamiltonian
\begin{equation}
H = \sum_{j=1}^L \left( S_j^x S_{j+1}^x + S_j^y S_{j+1}^y + h S_j^z \right),
\end{equation}
where $S_j^\alpha$ ($\alpha = x,y,z$) denote the spin-$1$ operators acting on the $j$-th lattice site, and $h$ is the strength of the uniform longitudinal field. Periodic boundary conditions are imposed such that $S_{L+1}^\alpha = S_1^\alpha$ for all spin components.

A prominent feature of this spin-$1$ XX chain is the presence of two distinct families of quantum many-body scar states \cite{Schecter:2019oej}. These are atypical eigenstates that violate the ETH and exhibit non-thermal dynamics, despite being embedded in an otherwise thermalizing many-body spectrum. They represent a paradigmatic example of weak ergodicity breaking, where ergodicity is lifted only for a measure-zero subset of eigenstates.

The first family is the bimagnon tower of scar states, constructed algebraically as
\begin{equation}
|S_n\rangle = \mathcal{N}(n) (J^+)^n |\Omega\rangle,
\end{equation}
for integer $n = 0,1,\dots,L$. The reference state $|\Omega\rangle = \bigotimes_{j=1}^L |m_j = -1\rangle$ is the fully polarized down state, where $m_j$ labels the local magnetization projection on site $j$. The normalization factor reads $\mathcal{N}(n) = \sqrt{\frac{(L-n)!}{n!\, L!}}$, and the collective raising operator $J^+$ is defined as $J^+ = \frac{1}{2}\sum_{j=1}^L (-1)^j (S_j^+)^2$, with $S_j^+ = S_j^x + \ii S_j^y$ being the standard local spin-$1$ raising operator. Each bimagnon state $|S_n\rangle$ is an exact eigenstate of the Hamiltonian with well-defined quantum numbers: energy $E_n = h(2n-L)$, total magnetization $m_n = 2n-L$, and crystal momentum $K_n = \frac{n\pi L}{2} \pmod{L}$. The second family, known as the bond-bimagnon tower, follows a similar algebraic construction; we refer the reader to the supplemental material of Ref.~\cite{Schecter:2019oej} for its explicit form.

\begin{figure}[tp]
  \centering
  \includegraphics[width=0.23\textwidth]{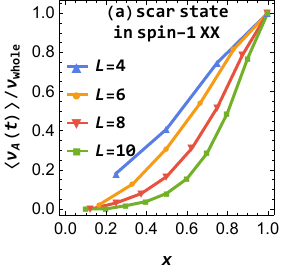}
  \includegraphics[width=0.23\textwidth]{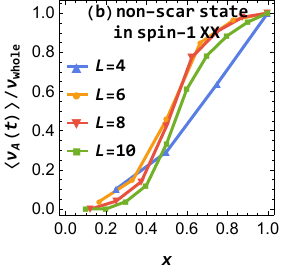}
  \caption{Rescaled time-averaged subsystem evolution speed after a quench from perturbed eigenstates in the spin-$1$ XX chain. \textit{Left}: Bimagnon scar state $|S_n\rangle$ with $n=3$. \textit{Right}: Non-scar eigenstate at the 1/4 position of the spectrum. We have taken $h=\sqrt{2}$. The time average is taken over the time window $t\in[2L,4L]$.}
  \label{FigureSpin1XX}
\end{figure}

We examine the dynamical signatures of these scar states by computing the rescaled time-averaged subsystem evolution speed after a quench from perturbed eigenstates, and contrast the behavior of scar states with that of generic thermal eigenstates. Examples of the results are presented in Fig.~\ref{FigureSpin1XX}. From the shape of the curves, we conclude that the bimagnon scar state clearly violates ETH, whereas the mid-spectrum non-scar eigenstate obeys ETH.

\section{Discussion}

In this work, we have proposed a perturbed eigenstate quench protocol as an efficient single-eigenstate diagnostic for the ETH, which avoids the computational overhead of explicitly constructing matching thermal ensembles. The method relies on a clear qualitative distinction in the profile of time-averaged subsystem evolution speed versus subsystem size ratio: ETH-obeying eigenstates feature an S-shaped curve with a distinct inflection point near half the system size, while ETH-violating states exhibit a convex J-shaped profile with no such transition. We have verified the reliability of this criterion across a series of canonical one-dimensional spin chain models spanning diverse dynamical regimes, including chaotic and integrable transverse-field Ising chains, disordered XXZ chains in both the chaotic and MBL phases, and a spin-1 XX chain with quantum many-body scar states. In all systems, our diagnostic correctly identifies thermalizing and non-thermalizing eigenstates, in full agreement with established theoretical expectations.

Nevertheless, several limitations of the present study should be noted. Our numerical simulations are based on exact diagonalization, which limits the accessible system sizes to relatively small values; a systematic finite-size scaling analysis is still needed to confirm the asymptotic behavior of the criterion in the thermodynamic limit. In addition, all demonstrations are carried out in one-dimensional lattice systems, and the generality of the evolution-speed signature in higher-dimensional lattices or continuum quantum many-body systems remains to be explored.

Looking ahead, the proposed framework opens up multiple avenues for further investigation. The protocol can be naturally generalized to a broader class of many-body systems, such as periodically driven Floquet systems and models with Hilbert-space fragmentation, where conventional ETH diagnostics face practical or conceptual difficulties. From an experimental standpoint, the quench scheme is compatible with existing quantum simulation platforms including cold-atom lattices and superconducting circuits, offering a feasible experimental probe of eigenstate thermalization at the single-eigenstate level.

\section*{Acknowledgements}

We would like to thank Markus Heyl, Reyhaneh Khasseh, and M.~A.~Rajabpour for helpful discussions and comments.
We acknowledge support from the National Natural Science Foundation of China (NSFC) grant number 12205217.
Numerical calculations for this study were performed on the high performance cluster at Center for Joint Quantum Studies (HPC-CJQS) of Tianjin University.


\begin{thebibliography}{10}

\bibitem{Deutsch:1991msp}
J.~M. Deutsch, \textit{{Quantum statistical mechanics in a closed system}},
  \href{http://dx.doi.org/10.1103/PhysRevA.43.2046}{Phys. Rev. A {\bfseries
  43}, 2046--2049 (1991)}.

\bibitem{Srednicki:1994mfb}
M.~Srednicki, \textit{{Chaos and Quantum Thermalization}},
  \href{http://dx.doi.org/10.1103/PhysRevE.50.888}{Phys. Rev. E {\bfseries 50},
  888--901 (1994)}, [\href{https://arxiv.org/abs/cond-mat/9403051}{{\ttfamily
  arXiv:cond-mat/9403051}}].

\bibitem{Rigol:2007mja}
M.~Rigol, V.~Dunjko and M.~Olshanii, \textit{Thermalization and its mechanism
  for generic isolated quantum systems},
  \href{http://dx.doi.org/10.1038/nature06838}{Nature {\bfseries 452}, 854--858
  (2008)}, [\href{https://arxiv.org/abs/0708.1324}{{\ttfamily
  arXiv:0708.1324}}].

\bibitem{Rigol:2006jrd}
M.~{Rigol}, V.~{Dunjko}, V.~{Yurovsky} and M.~{Olshanii}, \textit{{Relaxation
  in a Completely Integrable Many-Body Quantum System: An Ab Initio Study of
  the Dynamics of the Highly Excited States of 1D Lattice Hard-Core Bosons}},
  \href{http://dx.doi.org/10.1103/PhysRevLett.98.050405}{Phys. Rev. Lett.
  {\bfseries 98}, 050405 (2007)},
  [\href{https://arxiv.org/abs/cond-mat/0604476}{{\ttfamily
  arXiv:cond-mat/0604476}}].

\bibitem{Gornyi:2005mbj}
I.~V. {Gornyi}, A.~D. {Mirlin} and D.~G. {Polyakov}, \textit{{Interacting
  Electrons in Disordered Wires: Anderson Localization and Low-T Transport}},
  \href{http://dx.doi.org/10.1103/PhysRevLett.95.206603}{Phys. Rev. Lett.
  {\bfseries 95}, 206603 (Nov., 2005)},
  [\href{https://arxiv.org/abs/cond-mat/0506411}{{\ttfamily
  arXiv:cond-mat/0506411}}].

\bibitem{Basko:2006fmp}
D.~M. {Basko}, I.~L. {Aleiner} and B.~L. {Altshuler}, \textit{{Metal insulator
  transition in a weakly interacting many-electron system with localized
  single-particle states}},
  \href{http://dx.doi.org/10.1016/j.aop.2005.11.014}{Ann. Phys. {\bfseries
  321}, 1126--1205 (May, 2006)},
  [\href{https://arxiv.org/abs/cond-mat/0506617}{{\ttfamily
  arXiv:cond-mat/0506617}}].

\bibitem{Nandkishore:2014kca}
R.~Nandkishore and D.~A. Huse, \textit{{Many body localization and
  thermalization in quantum statistical mechanics}},
  \href{http://dx.doi.org/10.1146/annurev-conmatphys-031214-014726}{Ann. Rev.
  Condensed Matter Phys. {\bfseries 6}, 15--38 (2015)},
  [\href{https://arxiv.org/abs/1404.0686}{{\ttfamily arXiv:1404.0686}}].

\bibitem{Alet:2017dwx}
F.~{Alet} and N.~{Laflorencie}, \textit{{Many-body localization: An
  introduction and selected topics}},
  \href{http://dx.doi.org/10.1016/j.crhy.2018.03.003}{{C. R. Phys.} {\bfseries
  19}, 498--525 (2018)}, [\href{https://arxiv.org/abs/1711.03145}{{\ttfamily
  arXiv:1711.03145}}].

\bibitem{Abanin:2018yrt}
D.~A. Abanin, E.~Altman, I.~Bloch and M.~Serbyn,
  \textit{{$Colloquium$:~Many-body localization, thermalization, and
  entanglement}}, \href{http://dx.doi.org/10.1103/revmodphys.91.021001}{Rev.
  Mod. Phys. {\bfseries 91}, 021001 (2019)},
  [\href{https://arxiv.org/abs/1804.11065}{{\ttfamily arXiv:1804.11065}}].

\bibitem{Sierant:2024khi}
P.~Sierant, M.~Lewenstein, A.~Scardicchio, L.~Vidmar and J.~Zakrzewski,
  \textit{{Many-Body Localization in the Age of Classical Computing}},
  \href{https://arxiv.org/abs/2403.07111}{{\ttfamily arXiv:2403.07111}}.

\bibitem{Bernien:2017ubn}
H.~Bernien et~al., \textit{{Probing many-body dynamics on a 51-atom quantum
  simulator}}, \href{http://dx.doi.org/10.1038/nature24622}{Nature {\bfseries
  551}, 579--584 (2017)}, [\href{https://arxiv.org/abs/1707.04344}{{\ttfamily
  arXiv:1707.04344}}].

\bibitem{Turner:2017fxc}
C.~J. Turner, A.~A. Michailidis, D.~A. Abanin, M.~Serbyn and Z.~Papic,
  \textit{{Weak ergodicity breaking from quantum many-body scars}},
  \href{http://dx.doi.org/10.1038/s41567-018-0137-5}{Nature Phys. {\bfseries
  14}, 745--749 (2018)}, [\href{https://arxiv.org/abs/1711.03528}{{\ttfamily
  arXiv:1711.03528}}].

\bibitem{Serbyn:2020wys}
M.~Serbyn, D.~A. Abanin and Z.~Papi{\'c}, \textit{{Quantum many-body scars and
  weak breaking of ergodicity}},
  \href{http://dx.doi.org/10.1038/s41567-021-01230-2}{Nature Phys. {\bfseries
  17}, 675--685 (2021)}, [\href{https://arxiv.org/abs/2011.09486}{{\ttfamily
  arXiv:2011.09486}}].

\bibitem{Moudgalya:2021xlu}
S.~Moudgalya, B.~A. Bernevig and N.~Regnault, \textit{{Quantum many-body scars
  and Hilbert space fragmentation: a review of exact results}},
  \href{http://dx.doi.org/10.1088/1361-6633/ac73a0}{Rept. Prog. Phys.
  {\bfseries 85}, 086501 (2022)},
  [\href{https://arxiv.org/abs/2109.00548}{{\ttfamily arXiv:2109.00548}}].

\bibitem{Chandran:2022jtd}
A.~Chandran, T.~Iadecola, V.~Khemani and R.~Moessner, \textit{{Quantum
  Many-Body Scars: A Quasiparticle Perspective}},
  \href{http://dx.doi.org/10.1146/annurev-conmatphys-031620-101617}{Ann. Rev.
  Condensed Matter Phys. {\bfseries 14}, 443--469 (2023)},
  [\href{https://arxiv.org/abs/2206.11528}{{\ttfamily arXiv:2206.11528}}].

\bibitem{Garrison:2015lva}
J.~R. Garrison and T.~Grover, \textit{{Does a single eigenstate encode the full
  Hamiltonian?}}, \href{http://dx.doi.org/10.1103/PhysRevX.8.021026}{Phys. Rev.
  X {\bfseries 8}, 021026 (2018)},
  [\href{https://arxiv.org/abs/1503.00729}{{\ttfamily arXiv:1503.00729}}].

\bibitem{Dymarsky:2016ntg}
A.~Dymarsky, N.~Lashkari and H.~Liu, \textit{{Subsystem eigenstate
  thermalization hypothesis}},
  \href{http://dx.doi.org/10.1103/PhysRevE.97.012140}{Phys. Rev. E {\bfseries
  97}, 012140 (2018)}, [\href{https://arxiv.org/abs/1611.08764}{{\ttfamily
  arXiv:1611.08764}}].

\bibitem{Khasseh:2023kxw}
R.~Khasseh, J.~Zhang, M.~Heyl and M.~A. Rajabpour, \textit{{Identifying Quantum
  Many-Body Integrability and Chaos Using Eigenstate Trace Distances}},
  \href{http://dx.doi.org/10.1103/PhysRevLett.131.216701}{Phys. Rev. Lett.
  {\bfseries 131}, 216701 (2023)},
  [\href{https://arxiv.org/abs/2301.13218}{{\ttfamily arXiv:2301.13218}}].

\bibitem{Zhang:2024qya}
J.~Zhang, M.~A. Rajabpour, M.~Heyl and R.~Khasseh, \textit{{Subsystem evolution
  speed as indicator of relaxation}},
  \href{http://dx.doi.org/10.1103/PhysRevB.111.L140410}{Phys. Rev. B {\bfseries
  111}, L140410 (2025)}, [\href{https://arxiv.org/abs/2410.17798}{{\ttfamily
  arXiv:2410.17798}}].

\bibitem{Nielsen:2010oan}
M.~A. Nielsen and I.~L. Chuang, \textit{{Quantum Computation and Quantum
  Information}}.
\newblock Cambridge University Press, Cambridge, UK, 10th anniversary~ed.,
  2010,
  \href{http://dx.doi.org/10.1017/CBO9780511976667}{10.1017/CBO9780511976667}.

\bibitem{Lieb:1961fr}
E.~H. Lieb, T.~Schultz and D.~Mattis, \textit{{Two soluble models of an
  antiferromagnetic chain}},
  \href{http://dx.doi.org/10.1016/0003-4916(61)90115-4}{Annals Phys. {\bfseries
  16}, 407 (1961)}.

\bibitem{Katsura:1962hqz}
S.~Katsura, \textit{{Statistical mechanics of the anisotropic linear Heisenberg
  model}}, \href{http://dx.doi.org/10.1103/PhysRev.127.1508}{{Phys. Rev.}
  {\bfseries 127}, 1508 (1962)}.

\bibitem{Pfeuty:1970ayt}
P.~Pfeuty, \textit{{The one-dimensional Ising model with a transverse field}},
  \href{http://dx.doi.org/10.1016/0003-4916(70)90270-8}{Annals Phys. {\bfseries
  57}, 79 (1970)}.

\bibitem{Calabrese:2011vdk}
P.~Calabrese, F.~H.~L. Essler and M.~Fagotti, \textit{{Quantum Quench in the
  Transverse Field Ising Chain}},
  \href{http://dx.doi.org/10.1103/PhysRevLett.106.227203}{Phys. Rev. Lett.
  {\bfseries 106}, 227203 (2011)},
  [\href{https://arxiv.org/abs/1104.0154}{{\ttfamily arXiv:1104.0154}}].

\bibitem{Cassidy:2011smq}
A.~C. {Cassidy}, C.~W. {Clark} and M.~{Rigol}, \textit{{Generalized
  Thermalization in an Integrable Lattice System}},
  \href{http://dx.doi.org/10.1103/PhysRevLett.106.140405}{Phys. Rev. Lett.
  {\bfseries 106}, 140405 (2011)},
  [\href{https://arxiv.org/abs/1008.4794}{{\ttfamily arXiv:1008.4794}}].

\bibitem{Banuls:2010fkc}
M.~C. {Ba{\~n}uls}, J.~I. {Cirac} and M.~B. {Hastings}, \textit{{Strong and
  Weak Thermalization of Infinite Nonintegrable Quantum Systems}},
  \href{http://dx.doi.org/10.1103/PhysRevLett.106.050405}{Phys. Rev. Lett.
  {\bfseries 106}, 050405 (2011)},
  [\href{https://arxiv.org/abs/1007.3957}{{\ttfamily arXiv:1007.3957}}].

\bibitem{Kim:2014jfl}
H.~Kim, T.~N. Ikeda and D.~A. Huse, \textit{{Testing whether all eigenstates
  obey the eigenstate thermalization hypothesis}},
  \href{http://dx.doi.org/10.1103/PhysRevE.90.052105}{Phys. Rev. E {\bfseries
  90}, 052105 (2014)}, [\href{https://arxiv.org/abs/1408.0535}{{\ttfamily
  arXiv:1408.0535}}].

\bibitem{Sandvik:2010lkj}
A.~W. Sandvik, \textit{{Computational Studies of Quantum Spin Systems}},
  \href{http://dx.doi.org/10.1063/1.3518900}{AIP Conf. Proc. {\bfseries 1297},
  135 (2010)}, [\href{https://arxiv.org/abs/1101.3281}{{\ttfamily
  arXiv:1101.3281}}].

\bibitem{Pal:2010ezp}
A.~{Pal} and D.~A. {Huse}, \textit{{Many-body localization phase transition}},
  \href{http://dx.doi.org/10.1103/PhysRevB.82.174411}{Phys. Rev. B {\bfseries
  82}, 174411 (2010)}, [\href{https://arxiv.org/abs/1010.1992}{{\ttfamily
  arXiv:1010.1992}}].

\bibitem{Luitz:2015cep}
D.~J. {Luitz}, N.~{Laflorencie} and F.~{Alet}, \textit{{Many-body localization
  edge in the random-field Heisenberg chain}},
  \href{http://dx.doi.org/10.1103/PhysRevB.91.081103}{Phys. Rev. B {\bfseries
  91}, 081103 (2015)}, [\href{https://arxiv.org/abs/1411.0660}{{\ttfamily
  arXiv:1411.0660}}].

\bibitem{Schecter:2019oej}
M.~Schecter and T.~Iadecola, \textit{{Weak Ergodicity Breaking and Quantum
  Many-Body Scars in Spin-1 XY Magnets}},
  \href{http://dx.doi.org/10.1103/PhysRevLett.123.147201}{Phys. Rev. Lett.
  {\bfseries 123}, 147201 (2019)},
  [\href{https://arxiv.org/abs/1906.10131}{{\ttfamily arXiv:1906.10131}}].

\end{thebibliography}

\providecommand{\href}[2]{#2}\begingroup\raggedright\endgroup

\end{document}